\documentclass[11pt]{amsart}
\usepackage{amsbsy,amscd,amsfonts,amssymb,amstext,amsmath,amsthm,mathrsfs,amssymb}
\usepackage{graphicx}
\usepackage[below]{placeins}
\usepackage{color}
\usepackage{verbatim}

\begin{document}

\title[Gillespie algorithm with random rate constants]{Stochastic simulation of biochemical systems with randomly fluctuating rate constants}
\author{Chia Ying Lee}
\thanks{Statistical and Applied Mathematical Sciences Institute, 19 T.W. Alexander Drive, P.O. Box 14006, Research Triangle Park, NC 27709, USA}
\date{\today}
\maketitle

\begin{abstract}
In an experimental study of single enzyme reactions, it has been proposed that the rate constants of the enzymatic reactions fluctuate randomly, according to a given distribution. To quantify the uncertainty arising from random rate constants, it is necessary to investigate how one can simulate such a biochemical system. To do this, we will take the Gillespie's stochastic simulation algorithm for simulating the evolution of the state of a chemical system, and study a modification of the algorithm that incorporates the random rate constants. In addition to simulating the waiting time of each reaction step, the modified algorithm also involves simulating the random fluctuation of the rate constant at each reaction time. We consider the modified algorithm in a general framework, then specialize it to two contrasting physical models, one in which the fluctuations occur on a much faster time scale than the reaction step, and the other in which the fluctuations occur much more slowly. The latter case was applied to the single enzyme reaction system, using in part the Metropolis-Hastings algorithm to enact the given distribution on the random rate constants. The modified algorithm is shown to produce simulation outputs that are corroborated by the experimental results. It is hoped that this modified algorithm can subsequently be used as a tool for the estimation or calibration of parameters in the system using experimental data.
\end{abstract}

\section{Introduction}

Gillespie's Stochastic Simulation Algorithm has recently gained popularly as a method for simulating the evolution of biochemical systems.
Its advantage lies in its ability to capture the inherent stochasticity present in a chemical reaction system, and provide a full statistical description of the evolution of the system --- a point not addressed by traditional mass action theory, which, through a mathematical model of ordinary differential equations (ODEs), is able to capture only ensemble averaged behaviour and assumes a continuum of reactant concentrations.
Despite the fact that the deterministic mass action theory and the stochastic model of Gillespie's algorithm are equivalent in the limit of large system sizes (both assuming the well-mixed assumption), it is widely accepted that the stochastic model is more appropriate for biochemical applications, for the reason that biochemical systems commonly involve very low numbers of reactant molecules.
\begin{comment}
Basal levels of proteins within a cell exist in the hundreds {\color{red}[ref]}, whilst transcriptional dynamics in gene regulatory networks involves only one DNA and a handful of RNA molecules {\color{red}[ref]}.
\end{comment}
However, recent studies in the phenomenon of dynamic disorder of biomolecules reveal further stochasticity in certain biochemical systems that has yet to be accounted for by either of the two models: biochemical systems with randomly fluctuating rate constants.

Dynamic disorder refers to the fluctuation of the conformational state of a biomolecule, which may be attributed to the minimization of energy landscapes \cite{MinXieBag08}. 
These fluctuations are also associated with causing the fluctuation of the reaction rate constant of the biomolecule due to its changed conformational states \cite{Aus_etal75,LuXie_etal98}.
The effects of fluctuations due to dynamic disorder has been actively investigated by experimentalists \cite{Xie,YuXie_etal06,YangCao02,LuXie_etal98,MinXie_etal05,Zwanzig}, whilst models to describe this phenomenon, both on the molecular level and on the macroscopic kinetics level, have also been studied \cite{KouXie04,MinXie10,QianEls02,Xie2005,Xie2001,ChaIgo09,ChaIgo10,MofCheBus10}.
It is interesting to note that a subdiffusion of the conformational fluctuations based on fractional Brownian motion model was proposed by \cite{KouXie04}, in which a key assertion was the molecule's long-range memory of its conformational fluctuations resulting in fluctuations on a broad range of time scales.
In all the findings, it is widely realized that the effects of dynamic disorder on the reaction kinetics of biochemical reaction systems differ from the behaviour of systems where dynamic disorder is absent, and are not detectable by ensemble experiments or models.
One case in point is shown in \cite{Xie}, where experiments on the $\beta$-galactosidase enzyme reaction have uncovered interesting statistical properties --- heavy-tailed behaviour and correlation in the product formation rates --- which are not predicted by models with non-random rate constants. 
In particular, careful statistical analysis in \cite{Xie} lead the authors to propose that the rate constants in this enzyme reaction system take on a continuum of values and has a stationary gamma distribution.

In light of the findings in the single enzyme reaction, the purpose of this paper is to propose a modification of Gillespie's algorithm to allow for random rate constants to be incorporated into the simulation.
Whilst we retain the idea of the original algorithm, which is to simulate a sample trajectory for the time evolution of the system by simulating the successive occurrences of every reaction in the system, we extend the underlying stochastic model to account for both the simulation of the waiting time between reactions, as well as the simulation of the time evolution of the fluctuating rate constant. 
In essence, the stochastic model for the evolution of the chemical system is a joint distribution $(\tau,c)$ that models the interdependency between the waiting time $\tau$ and the changing rate constant $c$.
The joint distribution is a matter of modelling, thus would of course depend on the physical properties of the reactant molecules.
However, what we propose here is a general framework that forms a starting point for how one might go about the modification of the Gillespie's algorithm.
Subsequently, we specialize the general framework to the single enzyme example investigated in \cite{Xie}, and illustrate a successful way to incorporate physical constraints on the slowly interconverting conformers into model for the joint distribution.
It is here that the basic idea of Markov chain Monte Carlo is applied as a model to enforce the physical constraints, and numerical simulations from the algorithm thus derived are in good concert with the experimental results.

Before proceeding, we note how this paper differs from other works relating to the Gillespie's algorithm. 
One major limitation of the algorithm is its slowness, thus one broad area of research focuses on speeding up the algorithm.
Examples of these methods include the $\tau$-leaping method \cite{Gil01}, used to simulate more than one reaction at once, and the quasi-SSA and multiscale techniques, used for simulating stiff systems that possess reactions that occur on multiple time scales \cite{RatGil_etal03,CaoGilPet05,SamVla05,SinHenNem09}. 
Simulations can also fully or partially employ the use of the equivalent chemical master equation \cite{Mac_etal08,ChoiTar_etal10}.
Such speed-ups are not intended to change the fundamental stochastic properties of the chemical system, but typically to merely make approximations to the stochastic model or its chemical master equation to achieve better computational efficiency.
Other works have studied more fundamental modifications of the system model, such as fluctuating rate constants of white noise type \cite{SriTar_etal07,Xie2005}, or biochemical systems possessing rate constants that vary with time as a result of external factors such as cell growth or temperature changes \cite{Lu_etal,And07,Pao06}.
However, to the best of our knowledge, no works have addressed the general issue of randomly fluctuating rate constants, particularly the kind exhibited in the presence of dynamic disorder where the rate constants vary over a wide range of time scales comparable to or slower than the time scale of the reactions. 
Finally, we remark that there seems to be no easy way to reduce the problem of dynamic disorder into the original framework of Gillespie's algorithm; even if the conformational changes were to be represented as a finite number of basic reactions in the chemical system, such a reduction is inadequate for reproducing the full dynamics of a system that actually has a continuous range of conformational states and rate constants (see \cite{Xie} and its accompanying supplementary material).

\section{Deriving the modified SSA}\label{Derivations}

Before launching into the modifications of the Gillespie algorithm, we briefly review the ideas in the original algorithm. 
The Gillespie algorithm is derived from a probabilistic model of chemical reactions at the level of molecular interactions.
Given a system of chemical reactions $R_\mu$, indexed by $\mu$, of the form
\begin{equation*}
  \eta_1 E_1 + \dots + \eta_r E_r \stackrel{c_\mu}{\longrightarrow} \zeta_1 P_1 + \dots + \zeta_p P_p
\end{equation*}
we associate with each reaction $R_\mu$ a constant $c_\mu$ with the property that \\
\begin{equation*}
  \begin{array}{rcl}
    c_\mu \delta t & = & %
     \text{average probability that a particular combination of} \\
 & & \text{$R_\mu$ reactant molecules will react accordingly in the next} \\ 
 & & \text{infinitesimal time interval $\delta t$.}
  \end{array}
\end{equation*}
The constant $c_\mu$ takes the interpretation in the SSA as the rate constant of the chemical reaction%
\footnote{The rate constants for the SSA are directly related to those in the mass action kinetics, up to an appropriate scaling factor \cite{Gillespie SSA}.}.
Denote by $h_\mu$ the stoichiometric coefficient representing the total number of possible combinations of $R_\mu$ reactant molecules available. 
The \emph{propensity} $a_\mu := h_\mu c_\mu$ describes the rate for reaction $R_\mu$ to occur, in the sense that $a_\mu \delta t$ is the probability that reaction $R_\mu$ will occur in the next infinitesimal time interval $\delta t$.
It is then shown that the waiting time $\tau$ for the next reaction in the system to occur is exponentially distributed with rate $\sum_\mu a_\mu$ and density 
\begin{equation}\label{Eqn: orig SSA tau density}
f_{\tau} (\tau) = \left(\sum_\mu a_\mu\right)e^{-\sum_\mu a_\mu \tau}.
\end{equation}
Additionally, $a_\mu/(\sum_\mu a_\mu)$ is the probability that the reaction that occurred is $R_\mu$. 

The SSA produces a sample trajectory starting from an initialized the reactant state space at time $t=0$, followed by iterative simulation of subsequent reaction steps. 
Each reaction step is simulated by generating a random number $\tau$ from the density \eqref{Eqn: orig SSA tau density} and a random index $\mu$ for the reaction $R_\mu$, then updating the reactant state space and time step accordingly.

\subsection{A framework for random rate constants in Gillespie's algorithm}
Having been motivated to accommodate random rate constants into the SSA, we present a framework that allows us to extend the SSA to situations where the rate constant is a random variable. 
For ease of presentation, we first consider a single reaction scheme involving only one reactant. Subsequently, extending the derivation to reactions with more than one reactant, or to a system of reactions is straightforward.
Thus, consider
\begin{equation}\label{Rxn: fission eqn}
E \stackrel{c}{\longrightarrow} E+P
\end{equation}
The rate constant $c=c(t)$ is a continuous-time Markov process, which represents the time evolution of the fluctuating rate constant.
Further properties of the process $c(t)$ can be any physically relevant assumptions.
In this paper, we are primarily interested to have $c(t)$ possess a stationary distribution with density given by $w(c)$. 
The stationarity of $c(t)$ is a natural assumption in view of analogous steady state assumptions on the conformational state of a biomolecule \cite{ChaIgo10}.
The choice for the stationary density is a modelling issue that depends upon the reaction in question, or may be suggested from experimental data. For example, the gamma distribution was proposed in \cite{Xie},
\begin{equation}\label{Eqn: gamma density function}
w(c) = \frac{1}{b^a \Gamma (a)} c^{a-1} e^{-\frac{c}{b}}
\end{equation}
with parameters $a,b>0$ and where $\Gamma$ is the gamma function. 

Although the quantity $c$ is a randomly fluctuating process and is no longer constant, we will continue to use the term `rate constant' to refer to values that $c(t)$ takes at a given time $t$. We will also refer to the distribution of $c(t)$ as the `rate distribution'.

The starting point for the modification of the SSA is the idea that, given the rate constant at the time of the last reaction to be $c_0$, we want to derive a model for the joint density $f(\tau,c_1|c_0)$ of the waiting time $\tau$ for the next reaction to occur and the rate constant $c_1$ at the time when the next reaction occurs. 
Upon elucidating a model for $f(\tau,c_1|c_0)$, the algorithm proceeds similarly to the original SSA in a reaction-stepwise fashion, except that now a pair of random numbers $(\tau,c_1)$ is drawn according to $f(\tau,c_1|c_0)$, and both the time step and rate constant must be updated.
In this way, the algorithm sees only the rate constants $c_i$ at reaction times, and ignores any underlying properties of the underlying process $c(t)$.%
\footnote{By writing $f(\tau,c_1|c_0)$, we are implicitly assuming the Markovian property for both the reaction system and the process $c(t)$. The former case can be justified if we assume a well-mixed reaction system, in the sense that each reacting particle's location is uniformly distributed within the reaction volume. The latter case is a modelling assumption that is made in this paper for simplicity and illustrative purposes. Under other compelling physical motivation, one may be compelled to develop a non-Markovian model for the $c_i$s.}

Determining the joint density $f(\tau,c_1|c_0)$ is an issue of modelling, and should based on properties of the reactants and chemical system.
In general settings, using a joint distribution is a central aspect of the modified algorithm because it captures a wide range of cases where $\tau$ and $c_1$ may or may not be correlated.
One case is a scenario where the process $c(t)$ is modelled to evolve as an independent variable on which the distribution of $\tau$ depends as a dependent variable; the other end of the spectrum are situations where $(\tau,c_1)$ are independent, given $c_0$.
However, a more delicate and interesting structure of interdependence between $\tau$ and $c_1$ could arise in scenarios where, on the one hand the fluctuations of $c_1$ depend on the length of waiting time allotted, while on the other hand the waiting time depends on the dynamics of the rate constant.

To make the problem of modelling the joint density more tractable, it is convenient to factorize the joint density $f(\tau,c_1|c_0)$ into conditional densities,
\begin{equation}\label{conditional joint distri}
f_{\tau,c_1|c_0} = f_{c_1|c_0} \cdot f_{\tau|c_1,c_0} = f_{\tau|c_0} \cdot f_{c_1|\tau,c_0}.
\end{equation}
Here, we use $f$ as a generic notation for a density.
Either the first or second equality may be used for constructing the model.
Suppose we are given a model for $f_{\tau|c_0}$ and $f_{c_1|\tau,c_0}$. 
The algorithm then proceeds reaction-stepwise as shown in the following table.
The algorithm works analogously if we know $f_{c_1|c_0}$ and $f_{\tau|c_1,c_0}$ instead.
\vskip12pt
\fbox{\begin{minipage}{.9\linewidth}
Suppose $c_0$ is the rate constant at the most recent reaction. To find the waiting time and rate constant $(\tau,c)$ of the next reaction,
\begin{enumerate}
\item Pick $\tau$ randomly according to $f_{\tau|c_0}$.
\item Given $\tau$ and $c_0$, pick $c_1$ randomly according to $f_{c_1|\tau,c_0}$.
\end{enumerate}
Update state space and rate constant $c_1$. Progress the reaction time by $\tau$.
\end{minipage}}
\vskip12pt

\subsection{Formulating the modified Gillespie's algorithm}\label{c dynamics}

In order to use the modification of the Gillespie's algorithm, one has to first construct a model based on one of the two conditional density factorizations in \eqref{conditional joint distri}. 
In this paper, we will focus on the latter factorization.
Thus, we consider the conditional density $f_{c_1|\tau,c_0}(c_1|\tau,c_0)$, a.k.a. the transition kernel of $c(t)$, to be a model for how $c_1$ fluctuates over the waiting time interval of length $\tau$, given that it starts at $c_0$. In general, the transition kernel $f_{c_1|\tau,c_0}(c_1|\tau,c_0)$ should represent any physically relevant description of the $c$-dynamics.
Such a description should be constructed to incorporate information about the stationary distribution, as well as the physical constraints on the dynamics of $c$, that for instance could arise systems where reactants exhibit slow interconversion between conformational states (e.g. \cite{Xie}), or in systems with rapid fluctuation of rate constants (e.g. \cite{Xie2005}). 
In Section \ref{Section: Single enzyme reaction}, we will show a case study of slowly interconverting conformers in which we construct the transition kernel with help from the Metropolis-Hastings algorithm.
However, for this section, we will assume that $f_{c_1|\tau,c_0}(c_1|\tau,c_0)$ is given to us.

Suppose $f_{c_1|\tau,c_0}(c_1|\tau,c_0)$ has been determined. We use the second equality in equation \eqref{conditional joint distri} to obtain
\begin{align}\label{Eqn: 2nd derivation}
f_{\tau,c_1|c_0}(\tau,c_1|c_0) 
&= f_{c_1|\tau,c_0}(c_1|\tau,c_0) \cdot f_{\tau|c_0}(\tau|c_0) \nonumber\\
&= f_{c_1|\tau,c_0}(c_1|\tau,c_0) \cdot h\varphi_{c_0}(\tau) \cdot e^{-h\int_0^{\tau} \varphi_{c_0}(\tau') {\rm d}\tau'}
\end{align}
where $\varphi_{c_0}(\tau) = \mathbb{E}_{c_1}(c_1|\tau,c_0)$ is the conditional expected value of $c_1$ given $(\tau,c_0)$, and $h$ is the stoichiometric number associated with the number of reactant molecules.
In particular, we have derived in Appendix \ref{Appdx: tau, c0} that
\begin{equation}
  f_{\tau|c_0}(\tau|c_0) = h\varphi_{c_0}(\tau) e^{-h\int_0^{\tau} \varphi_{c_0}(\tau') d\tau'}.
\end{equation} 
This formula indicates that the effective propensity of the reaction is $a_\mu (\tau) = h\varphi_{c_0}(\tau)$, in the sense that $\int_{t+\tau}^{t+\tau+\delta\tau} a_{\mu}(\tau') d\tau'$ is the probability that the reaction occurs in the infinitesimal time interval $[t+\tau, t+\tau+\delta\tau)$.

We observe that the distribution of $\tau$ depends only on the transition dynamics of the process $c(t)$ --- specifically, it depends only on the \emph{conditional mean} of $c(t)$ in the waiting time interval.
This is an important observation, because it simplifies the modelling demands to rest only on providing a model of $f_{c_1|\tau,c_0}$.
Once $f_{c_1|\tau,c_0}$ is determined, the model for $f_{\tau|c_0}$ is automatically available, thereby reducing the complexity of the model.

Since the cumulative distribution function of $f_{\tau|c_0}$ is
\begin{equation}\label{Eqn: 2nd derivation marginal cdf}
F(\tau|c_0) = 1-e^{-h\int_0^{\tau} \varphi_{c_0}(\tau') d\tau'},
\end{equation}
sampling from the density $f_{\tau|c_0}(\tau|c_0)$ can be achieved by inverting the expression
\begin{equation*}
  \int_0^\tau h \varphi_{c_0}(\tau') d\tau' = -\log r
\end{equation*}
for $\tau$, where $r$ is a uniform random variable on $[0,1]$.
In other words, defining $\Phi_{c_0}(\tau) = \int_0^{\tau} \varphi_{c_0}(\tau') d\tau'$, we find $\tau$ by the transformation 
\begin{equation*}
  \tau = \Phi_{c_0}^{-1}(-\frac{1}{h}\log (1-r)).
\end{equation*}
However, a closed form formula for $\Phi_{c_0}^{-1}$ may not always be readily available, except for special forms of the function $\varphi_{c_0}(\tau')$ (see e.g., \cite{Lu_etal}), and hence numerical approximation procedures will have to be applied to compute $\tau$. This may become computationally intensive, but we will not discuss these computational issues here.

\subsection{Some equivalent formulations}
\paragraph{Semi-Markovian approximations}\label{Semi-Markov}
Kou et al. \cite{Xie2005} described a so-called semi-Markovian approximation in which the fluctuation of the rate constant exhibits large variability from the time of one reaction to the next.
This is characterized by rapidly fluctuating dynamics of the rate constant in a shorter time scale than the reaction waiting times. 
Thus, the approximation aspect of this model assumes the process $c(t)$ to be characterized by infinitesimally small correlation lengths, so that the rate constant at the next reaction is independent of its value at the last reaction.
This approximation leads to setting $f_{c_1|\tau,c_0}(c_1|\tau,c_0) = w(c_1)$.
It is no surprise that the semi-Markovian approximation yields no correlation between successive reaction waiting times \cite{Xie2005}.

\paragraph{Time independent transition kernels.}
If $f_{c_1|\tau,c_0}$ does not depend on $\tau$, then we have $f_{c_1|\tau,c_0} (c_1|\tau,c_0) = f_{c_1|c_0} (c_1|c_0)$ and $f_{\tau|c_1,c_0} (\tau|c_1,c_0) = f_{\tau|c_0}(\tau|c_0) = h\varphi_{c_0} e^{-\tau h \varphi_{c_0}}$.
Then it is equivalent to use either the first or second equalities in \eqref{conditional joint distri}. 
However, in general cases, it is not trivial to derive a formula for $f_{\tau|c_1,c_0} (\tau|c_1,c_0)$.

\paragraph{Reduction to the original SSA.}
Equation \eqref{conditional joint distri} reduces to Gillespie's original SSA with a non-random rate constant $\bar{c}$ by the special choice of $w(c)=\delta(c-\bar{c})$, where $\delta$ represents the Dirac mass at 0. 
This leads to setting $f_{c_1|\tau,c_0} (c_1|\tau,\bar{c}) = \delta(c_1-\bar{c})$ in equation \eqref{Eqn: 2nd derivation}. 
It is clear that the algorithm will always pick $c_1 = \bar{c} = c_0$ almost surely.

\paragraph{Non-random but time varying rate constants.}
If the rate constant is a non-random function of time, $c(t)$, we can still apply the framework without the stationarity assumption. The conditioning will be on $c_0$ at the current time $t$, and thus we set $f_{c_1|\tau,t,c_0} (c_1|\tau,t,c(t)) = \delta(c_1-c(t+\tau))$. 
This case had been studied in \cite{Lu_etal,And07,Pao06}.

\section{Single enzyme reactions - an example}
\label{Section: Single enzyme reaction}

We illustrate an application of the modified Gillespie algorithm to the simulation of single enzyme reactions, en route proposing a method to construct a model for the transition kernel $f_{c_1|\tau,c_0}(c_1|\tau,c_0)$ using the Metropolis-Hastings algorithm. 
Here, we consider a simple version of the enzyme reaction, where an enzyme $E$ binds reversibly with the substrate $S$ to form a complex $ES$, which then dissociates to release the product $P$. 
The reaction is given by the typified enzyme kinetics scheme 
\begin{equation}\label{Rxn: enzyme}
E+S {{k_1 \atop\longrightarrow}\atop{\longleftarrow\atop k_{-1} }} ES \stackrel{c}{\longrightarrow} E + P
\end{equation}
where $k_1$ and $k_{-1}$ are non-random rate constants associated with the original Gillespie algorithm, for the complex formation and dissociation reactions, and $c$ is the random rate constant for the product formation reaction with density $w(c)$.

When $w(c)$ is a non-random constant $\bar{c}$, classical Michaelis-Menten kinetics provides a deterministic relation between the rate of product formation and the substrate concentration,
\begin{equation}\label{Eqn: classical Michaelis-Menten}
\frac{{\rm d}[P]}{{\rm d}t} = \frac{v_{{\rm max}}[S]}{[S]+K_M}
\end{equation}
where $v_{\max} = ([E]+[ES])\bar{c}$ is the maximum enzyme velocity and $K_M = \frac{k_{-1}+\bar{c}}{k_1}$ is the Michaelis constant. 
Here and in future, the notation $[E],[S]$, etc., denotes the concentration of the reactant species $E,S$, etc.
In the case of a single enzyme system, it is more appropriate when investigating the rate of formation of the product to consider the waiting time between the formation of successive products $P$, which we refer to as the turnover time $\tau$. 
The rate of product formation can be reformulated as
\begin{equation}\label{Eqn: single enzyme MM}
\frac{1}{\langle\tau\rangle} = \frac{\bar{c}[S]}{[S]+K_M}
\end{equation}
 where $\langle \tau \rangle$ is the ensemble average of $\tau$ (see \cite{QianEls02,Xie2001,Xie2005}).

When dynamic disorder is present, the rate constant $c$ fluctuates according to a distribution $w(c)$. In this case, the rate of product formation in single enzyme reactions under a quasi-static condition of dynamic disorder has been shown to take an analogous form \cite{Xie2005}
\begin{equation}\label{Eqn: dynamic disorder MM}
\frac{1}{\langle\tau\rangle} = \frac{\chi[S]}{[S]+C_M}
\end{equation}
where 
\begin{equation*}
  \chi = \frac{1}{\int_0^\infty\frac{w(c)}{c}{\rm d}c}, \qquad \text{and} \qquad C_M = \frac{(k_{-1}+\chi)}{k_1}
\end{equation*}
are the harmonic mean of $c$ and the effective Michaelis constant, respectively.

In the next section, we derive an algorithm to simulate the single enzyme reaction system using the method of modifying the SSA as described in the previous section. In addition to making a simple extension to multiple reactions, we propose a method to model the joint distribution $f(\tau,c_1)$ that can be applied quite generally. 
A major consideration in the modelling comes from the paper by English et al \cite{Xie}, in which it was proposed that the single enzyme reaction exhibits dynamic disorder with the rate constant $c$ distributed according to the gamme distribution \eqref{Eqn: gamma density function}. 
More importantly, the dynamic disorder was suggested to be a result of conformational fluctuations of the enzyme that occur at time scales much larger than product formation time lengths.

\subsection{Modelling slowly interconverting conformers} \label{Section: Slow interconversion}

We focus our attention for the time being on the product formation reaction step, $ES \stackrel{w(c)}{\longrightarrow} P+E$. 
Equation (\ref{Eqn: dynamic disorder MM}) holds with the assumption that interconversion of conformers occur much slower than the reaction. 
This assumption is incorporated into the modified SSA by restricting the rate constant at the next time step, $c_1$, to change by only a small amount between each reaction.
Specifically, we use the conditional density as described in \S\ref{c dynamics}, and model $f_{c_1|\tau,c_0}(c_1|\tau,c_0)$ to be a transition kernel whose support lies in an $\varepsilon$-small interval around $c_0$. 
As a first pass, we also make a simplifying assumption that $f_{c_1|\tau,c_0}(c_1|\tau,c_0)$ does not depend much on $\tau$, and consider in its place a conditional density $f_{c_1|c_0}(c_1|c_0)$ whose support lies in a fixed $\varepsilon$-small interval around $c_0$.
Alternatively, removing the simplifying assumption can be done by modelling the $\varepsilon$-small interval to grow with $\tau$.

Upon deciding on the support interval for $f_{c_1|\tau,c_0}(c_1|\tau,c_0)$, we then make use of the Metropolis-Hastings algorithm to construct a Markov chain for the successive values of $c_1$, as follows.
Define the density
\begin{align} \label{Eqn: uniform g}
  g(c_1|c_0) = 
  \begin{cases}
    \frac{1}{\varepsilon}, &\text{if } c_1\in (c_0-\frac{\varepsilon}{2}, c_0+\frac{\varepsilon}{2}) \\
    0, &\text{otherwise}
  \end{cases}
\end{align}
and define the \emph{acceptance probability} 
\begin{align*}
  \alpha(c_1|c_0) = \min \left(1 ,\,\frac{w(c_1)}{w(c_0)} \frac{g(c_1|c_0)}{g(c_0|c_1)}\right) = \min \left(1 ,\,\frac{w(c_1)}{w(c_0)}\right)
\end{align*}
for $|c_1-c_0|< \varepsilon/2$.
Then the transition probability for $c_1$ does not depend on $\tau$ and is given by
\begin{align} \label{Eqn: c-dynamics for slow interconversion}
f_{c_1|c_0}(c_1|c_0) &= g(c_1|c_0)\alpha(c_1|c_0) \nonumber\\
&= g(c_1|c_0) \min \left(1 ,\,\frac{w(c_1)}{w(c_0)}\right). 
\end{align}

It is clear that the $c_1$ generated in this way will lie in the interval $(c_0 -\varepsilon/2, c_0 +\varepsilon/2)$, and it is easy to check that $f_{c_1|c_0}(c_1|c_0)$ is the transition kernel of a Markov chain with $w(c)$ as its unique stationary distribution. 
That is, if we pick a random number $r_c$ from the forcing distribution $g(c_1|c_0)$, and accept it with probability $\alpha(r_c|c_0)= \min \left(1 ,\, \frac{w(r_c)}{w(c_0)} \frac{g(r_c|c_0)}{g(c_0|r_c)}\right)$, we get that $r_c$ satisfies the transition kernel $f_{c_1|\tau,c_0}(c_1|\tau,c_0)$, and moreover $r_c$ is $w(c)$-distribution provided $c_0$ is $w(c)$-distributed.

From \eqref{Eqn: 2nd derivation}, we automatically have a formula for the waiting time distribution
\begin{equation*}
  f_{\tau|c_0}(\tau|c_0) = h_{ES} \varphi_{c_0} e^{-h_{ES} \int_0^\tau \varphi_{c_0} d\tau'}
\end{equation*}
where $h_{ES}$, the stoichiometric number of the complex $ES$, is 1 if the enzyme is in complex form $ES$ and 0 if it is in free form $E$, and
\begin{equation*}
  \varphi_{c_0} = \int_0^\infty c_1 f_{c_1|\tau,c_0}(c_1|\tau,c_0) d c_1 = \int_{c_0 -\varepsilon/2}^{c_0 +\varepsilon/2} \frac{c_1}{\varepsilon} \min \left(1 ,\,\frac{w(c_1)}{w(c_0)}\right) \,d c_1.
\end{equation*}
Thus, the effective propensity of the product formation step is $h \varphi_{c_0}$.

Before moving on, it should be remarked that the use of the Metropolis-Hastings algorithm is solely as a way to construct a Markov chain possessing a stationary distribution. Although its original development was to simulate samples from a stationary distribution that is otherwise difficult to simulate from, this purpose plays no role in its application here. Quite the contrary, the stationary distribution may be an easily simulated distribution, as is the case in our example with the gamma distribution. Rather, any Markov chain can be used as a model, so long as it satisfies the physical constraints --- a maximum range of fluctuations and a stationary distribution. The use of the Metropolis-Hasting algorithm so happens to be a convenient choice because it provides an easy way to specify the maximum range of fluctuations via the density $g$ whilst maintaining the stationarity due to the acceptance probability $\alpha$. And we will see in the simulation results that this modelling choice corroborates the experimental results.

\subsection{The modified SSA for single enzyme reactions} 
\label{Section: Modified SSA for single enzyme reactions}

\begin{figure}
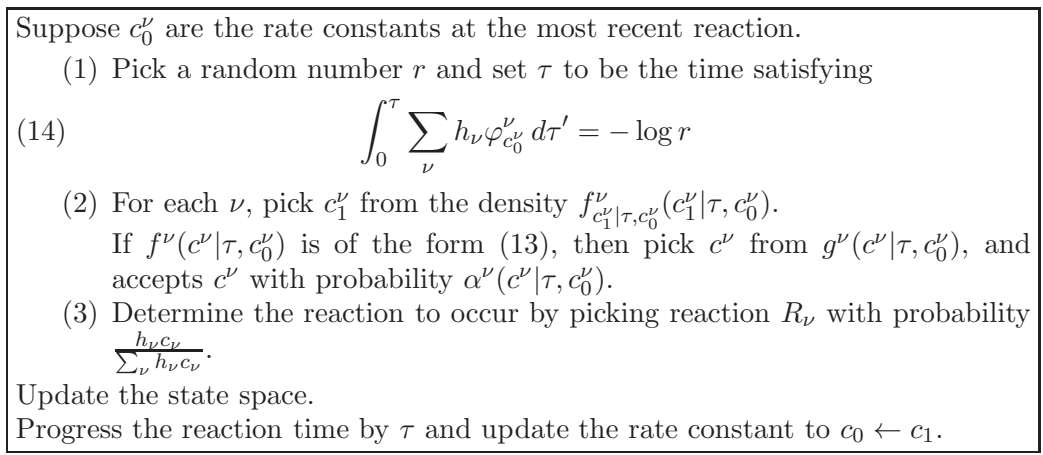

\centering
\fbox{\begin{minipage}{.9\linewidth}
Suppose $c_0^\nu$ are the rate constants at the most recent reaction.
\begin{enumerate}
\item Pick a random number $r$ and set $\tau$ to be the time satisfying \begin{equation} \label{Eqn: waiting time in system}
  \int_0^\tau \sum_{\nu} h_\nu \varphi_{c_0^\nu}^\nu \,d\tau' = -\log r
\end{equation}
\item For each $\nu$, pick $c_1^\nu$ from the density $f^\nu_{c_1^\nu|\tau,c_0^\nu}(c_1^\nu|\tau,c_0^\nu)$.\\
  If $f^\nu(c^\nu|\tau,c_0^\nu)$ is of the form \eqref{Eqn: c-dynamics for slow interconversion}, then pick $c^\nu$ from $g^\nu (c^\nu|\tau,c_0^\nu)$, and accepts $c^\nu$ with probability $\alpha^\nu (c^\nu|\tau,c_0^\nu)$.
\item Determine the reaction to occur by picking reaction $R_\nu$ with probability $\frac{h_\nu c_\nu}{\sum_{\nu} h_\nu c_\nu}$.
\end{enumerate}
Update the state space. \\
Progress the reaction time by $\tau$ and update the rate constant to $c_0 \leftarrow c_1$.
\end{minipage}}
\caption{Modified Gillespie Algorithm}
\label{Algo: Slow interconversion}
\end{figure}

Finally, the Gillespie algorithm is adapted to integrate the procedure from the previous subsection into the algorithm for the entire enzyme reaction system which involves more than one reaction. 
In this reaction system, let $\#S$ be the number of substrate molecules, and $h_{E}=1-h_{ES}$. 
The propensities of the reversible enzyme-substrate complex formation reaction are $h_{E}(\#S)k_1$, $h_{ES}k_{-1}$ and, given the current rate constant $c_0$, the propensity of the product-forming reaction is $h_{ES} \varphi_{c_0}$.
The update quantities to be determined for each reaction step are the waiting time $\tau$, the new rate constant $c_1$ for the product formation reaction, as well as the reaction that occurs.

The waiting time $\tau$ to the next reaction is chosen to satisfy
\begin{equation}
  \int_0^\tau h_{ES} \varphi_{c_0}(\tau') + h_{E}(\#S) k_1 + h_{ES} k_{-1} \,d\tau' = -\log r
\end{equation}
for a uniform random number $r$. Note that here, due to our assumption that $g(c_1|c_0)$ is time-independent, $\varphi_{c_0}(\tau')$ is constant is $\tau'$ and thus the evaluation of $\tau$ is a direct algebraic evaluation.
Next, the new rate constant $c_1$ is chosen by picking $c$ from the density $g(c|c_0)$, and accepting $c$ with probability $\alpha(c|c_0)$. If it is accepted, set $c_1=c$, otherwise it is rejected and set $c_1=c_0$.
Finally, the reaction that occurs is chosen in a standard way of comparing the relative propensities of the reactions \emph{at the time of the next reaction}\cite{And07}: for the three reactions, complex formation, complex dissociation, and product formation reactions, the probabilities of their occurrence are proportional to
\begin{equation} \label{Eqn: propensities for enzyme reaction}
  h_{E} (\#S) k_1, \quad h_{ES} k_{-1}, \quad h_{ES} c_1,
\end{equation}
respectively.
This leads to the modified SSA for simulating slowly interconverting conformers in single enzyme reactions.

For an arbitrary chemical system of reactions $R_{\nu}$ with random rate constants drawn from the distribution $f^{\nu}_{c_1^\nu|\tau,c_0^\tau}$ and effective propensities $h_{\nu} \varphi_{c_0^\nu}^{\nu}(\tau)$, the generalization of the modified Gillespie algorithm is obvious.
Fig. \ref{Algo: Slow interconversion} summarizes the algorithm for each iteration of the reaction step.

\subsection{Simulation results} \label{Section: Simulation results}

\begin{figure}
\centering
\includegraphics[width=\linewidth]{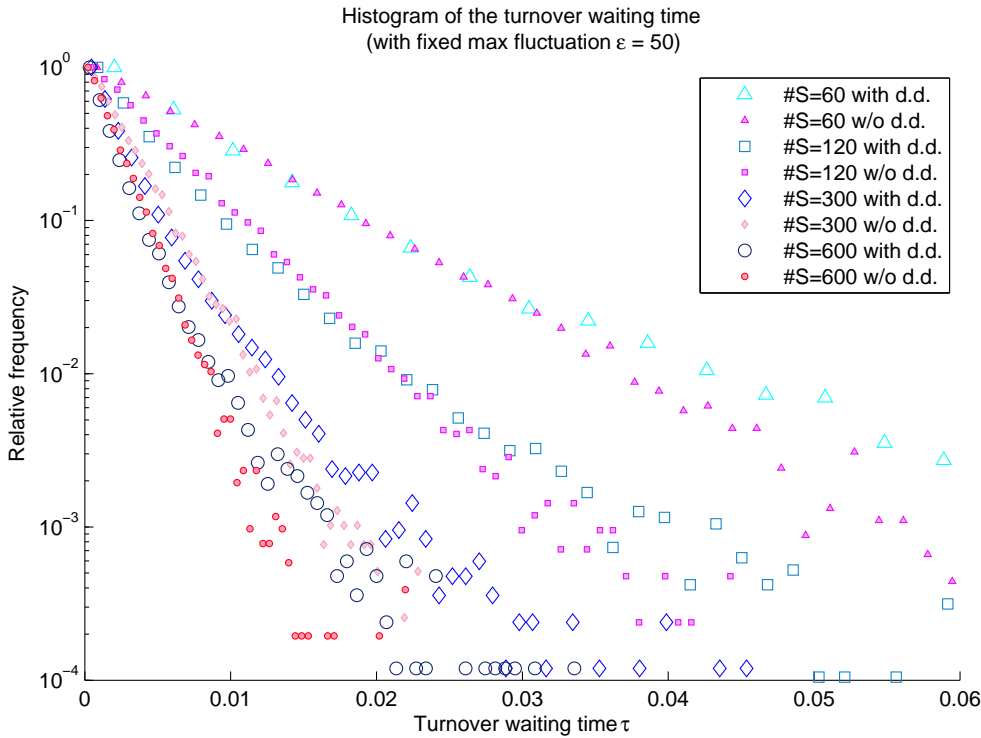}
\caption{Histogram of turnover waiting times, normalized so that the relative frequency of the smallest time bin equals 1. The heavy-tailed behaviour (large blue markers) due to dynamic disorder is juxtaposed against the exponentially light-tailed behavior in the absence of dynamic disorder (small pink markers).}
\label{Fig: histogram turnover fixepsi}
\end{figure}

To simulate a single enzyme system, we run the simulation starting with one enzyme molecule and $\#S=60,120,300,600$ number of free substrate molecules. 
We assume a buffered substrate solution, that is, $\#S$ remains constant even if a complex formation reaction occurs.
The rate constants were $k_1=50$, $k_{-1}=18300$, and $c$ follows a gamma distribution with parameters $a=4.2$, $b=220$ in equation \eqref{Eqn: gamma density function}. 
The fluctuation of $c$ is at most $\varepsilon = 50s^{-1}$ in \eqref{Eqn: uniform g}, which we fixed for all values of $\#S$.
With these parameters, each product formation occurs on average once in 30 complex dissociation reactions.

While the simplifying assumption is limited, the simulation results nonetheless show several features similar to those obtained experimentally by English et al., \cite{Xie}. 
One key feature is the heavy-tailed property of the distribution of the turnover time $\tau$, a phenomenon that is not observed if the rate constant is non-random. 
In the latter case, the distribution of the turnover time is known to exhibit exponential decay \cite{Xie2005}.
Figure \ref{Fig: histogram turnover fixepsi} shows the heavy-tailed property in a histogram of turnover times when the modified algorithm was used (dots), compared against the exponential decay obtained from the model without dynamic disorder (crosses) with a non-random rate constant $c \equiv \bar{c} = \mathbb{E}_{w}[c]$. 
For each value of $\#S$, the heavy-tailed property is most apparent the rare regime of large turnover times.

\begin{figure}
\centering
\includegraphics[width=.75\linewidth]{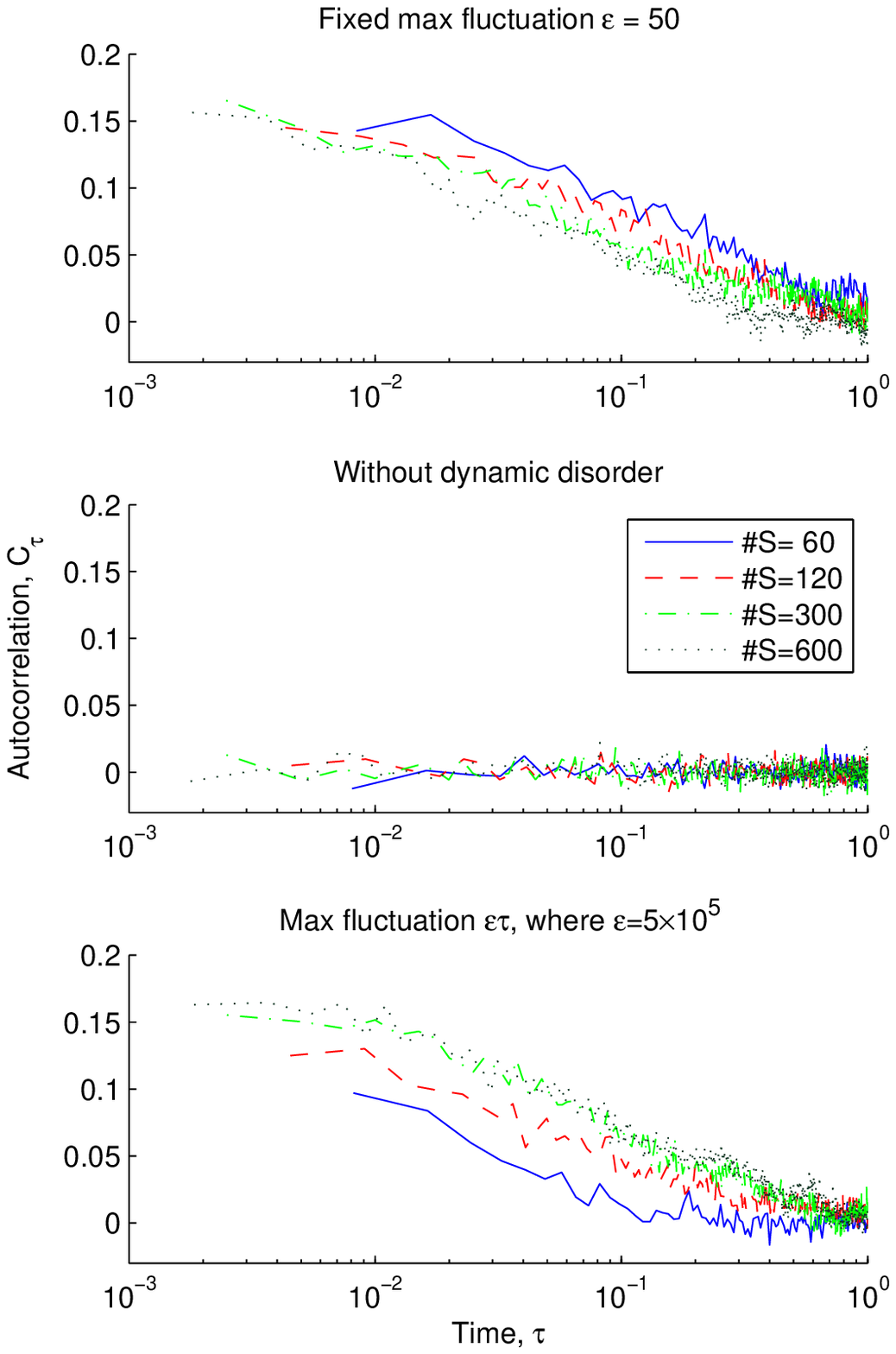}
\caption{Autocorrelation graph. The top panel shows the autocorrelation $C_\tau$ for the fixed $\varepsilon$ model in Section \ref{Section: Slow  interconversion}; the middle panel shows the autocorrelation for a model without dynamic disorder; the bottom panel shows the autocorrelation for the model with time-dependent $c$-dynamics in Section \ref{Section: Time dependent c-dynamics}. }
\label{Fig: autocorrelation}
\end{figure}

The crux of the work by English et al. is the discovery of correlations between successive turnover times of a single enzyme molecule, where they found that short turnover times are more likely to be followed by short turnover times, and vice versa. 
The top panel of Figure \ref{Fig: autocorrelation} shows the autocorrelation function of the turnover time series $\{\tau_m\}$, computed as
\begin{equation*}
  C_{\tau} (m) = \frac{\langle (\tau_m-\langle\tau\rangle) (\tau_0-\langle\tau\rangle) }{\langle (\tau-\langle\tau\rangle)^2 \rangle}
\end{equation*}
and converting $C_\tau(m)$ to $C_\tau(t)$ using $t = m\langle \tau\rangle$.
By construct, our model ensures that such positive correlations are produced, but that the systems containing different numbers of substrate molecules exhibit the same autocorrelation behavior shifted horizontally. 
This is attributable to the fact that the variance of the forcing distribution $g(c|c_0)$ governs the correlation between successive $c$, so that the same value of $\varepsilon$ applied for each value of $\#S$ necessarily results in the same autocorrelation behavior. 
This result differs from English's experimental result, which show an increasing degree of autocorrelation for increasing substrate concentrations. 
Heuristically, the experimental results can be rationalized by noting that, when $\#S$ is low, the complex-forming reaction becomes rate limiting, wider fluctuations of $c$ are attainable during the longer waiting time for the reaction, resulting in smaller correlation between turnover times.
This heuristic indicates that the fixed $\varepsilon$ model is inadequate.

\subsection{Modelling time dependence in the $c$-dynamics}
\label{Section: Time dependent c-dynamics}

It is fair to say that the simplified model in Section \ref{Section: Slow  interconversion} with fixed $\varepsilon$ is a poor model for the true dynamics of $c$.
To improve the model, it becomes necessary to incorporate the time dependence in the $c$-dynamics.
As a next step toward this goal, we consider a model in which the amount of fluctuation of the rate constant $c$ depends on the length of the reaction waiting time.
In place of \eqref{Eqn: uniform g}, we use an interval that increases linearly with time,
\begin{align} \label{Eqn: uniform g with tau}
  g(c_1|\tau,c_0) = 
  \begin{cases}
    \frac{1}{\tilde{\varepsilon}\tau}, &\text{if } c_1\in (c_0-\frac{\tilde{\varepsilon}\tau}{2}, c_0+\frac{\tilde{\varepsilon}\tau}{2}) \\
    0, &\text{otherwise}
\end{cases}
\end{align}
As usual, the acceptance probability is given by $\alpha(c_1|\tau,c_0) = \min \left(1, \frac{w(c_1) g(c_1|\tau,c_0)}{w(c_0) g(c_0|\tau,c_1)} \right)$ for $c_1 \in (c_0-\frac{\tilde{\varepsilon}\tau}{2}, c_0+\frac{\tilde{\varepsilon}\tau}{2})$, and the transition probability is $f_{c_1|\tau,c_0} = g(c_1|\tau,c_0) \alpha(c_1|\tau,c_0)$.
The waiting time distribution, given in expression \eqref{Eqn: 2nd derivation}, is
\begin{equation*}
  f_{\tau|c_0} (\tau|c_0) = h \varphi_{c_0}(\tau) e^{ -h \int_0^\tau \varphi_{c_0} (\tau') d\tau'}
\end{equation*} 
where the effective propensity $\varphi_{c_0}(\tau)$ now does depend on $\tau$,
\begin{equation*}
  \varphi_{c_0}(\tau) = \int_0^\infty c_1 f_{c_1|\tau,c_0} dc_1 = \int_{c_0 -\frac{\tilde{\varepsilon}\tau}{2}}^{c_0 +\frac{\tilde{\varepsilon}\tau}{2}} \frac{c_1}{\varepsilon\tau} \min \left(1, \frac{w(c_1)}{w(c_0)} \right) dc_1.
\end{equation*}
The algorithm in Diagram \ref{Algo: Slow interconversion} is then applied to the model.

One immediate difficulty with incorporating time-dependence into the Gillespie algorithm is that sampling the waiting time distribution involves inverting the equation \eqref{Eqn: waiting time in system}.
In general, $\varphi_{c_0}^\nu(\tau)$ may be a complicated function and an explicit expression for the waiting time $\tau$ is difficult to obtain.
Consequently it becomes necessary to solve for $\tau$ numerically, and this procedure becomes computationally intensive.
In our simulations, assuming that $\tau$ is sufficiently small, we solve for $\tau$ by linearizing the LHS of \eqref{Eqn: waiting time in system} around 0.
In this way, we speed up the algorithm albeit at the expense of admitting some error.

\begin{figure}
\centering
\includegraphics[width=\linewidth]{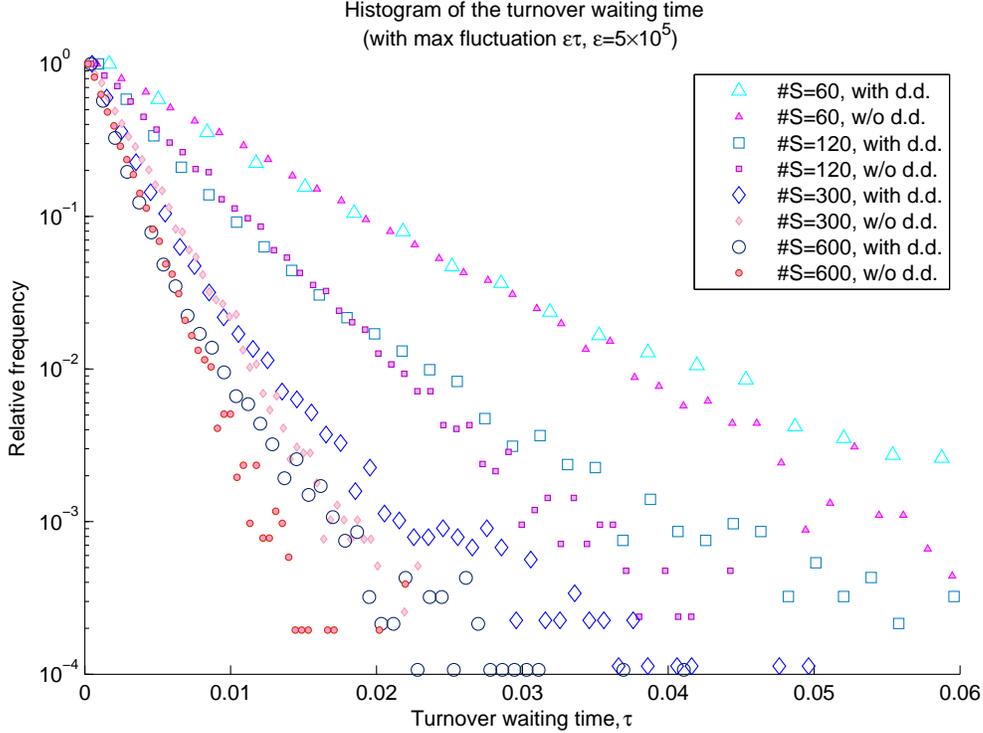}
\caption{Histogram of turnover waiting times, normalized so that the relative frequency of the smallest time bin equals 1. The time-dependent $c$-dynamics model (large blue markers) and the model without dynamic disorder (small pink markers) are shown.}
\label{Fig: histogram turnover epsitau}
\end{figure}

Figure \ref{Fig: histogram turnover epsitau} shows the histogram of the turnover times for increasing numbers of substrate molecules, $\# S = 60,120,300,600$. Here, we took $\tilde{\varepsilon} = 5\times10^5{\rm s}^{-2}$. Similar to Figure \ref{Fig: histogram turnover fixepsi}, the heavy tail behaviour is also observed. The key difference of the time-dependent model is seen in the autocorrelation graph in Figure \ref{Fig: autocorrelation}, where the autocorrelation increases with increasing $\#S$. Compared to the fixed $\varepsilon$ model, the autocorrelation behaviour is in closer agreement qualitatively with the experimental results of English et al.

\begin{figure}
\centering
\includegraphics[width=.75\linewidth]{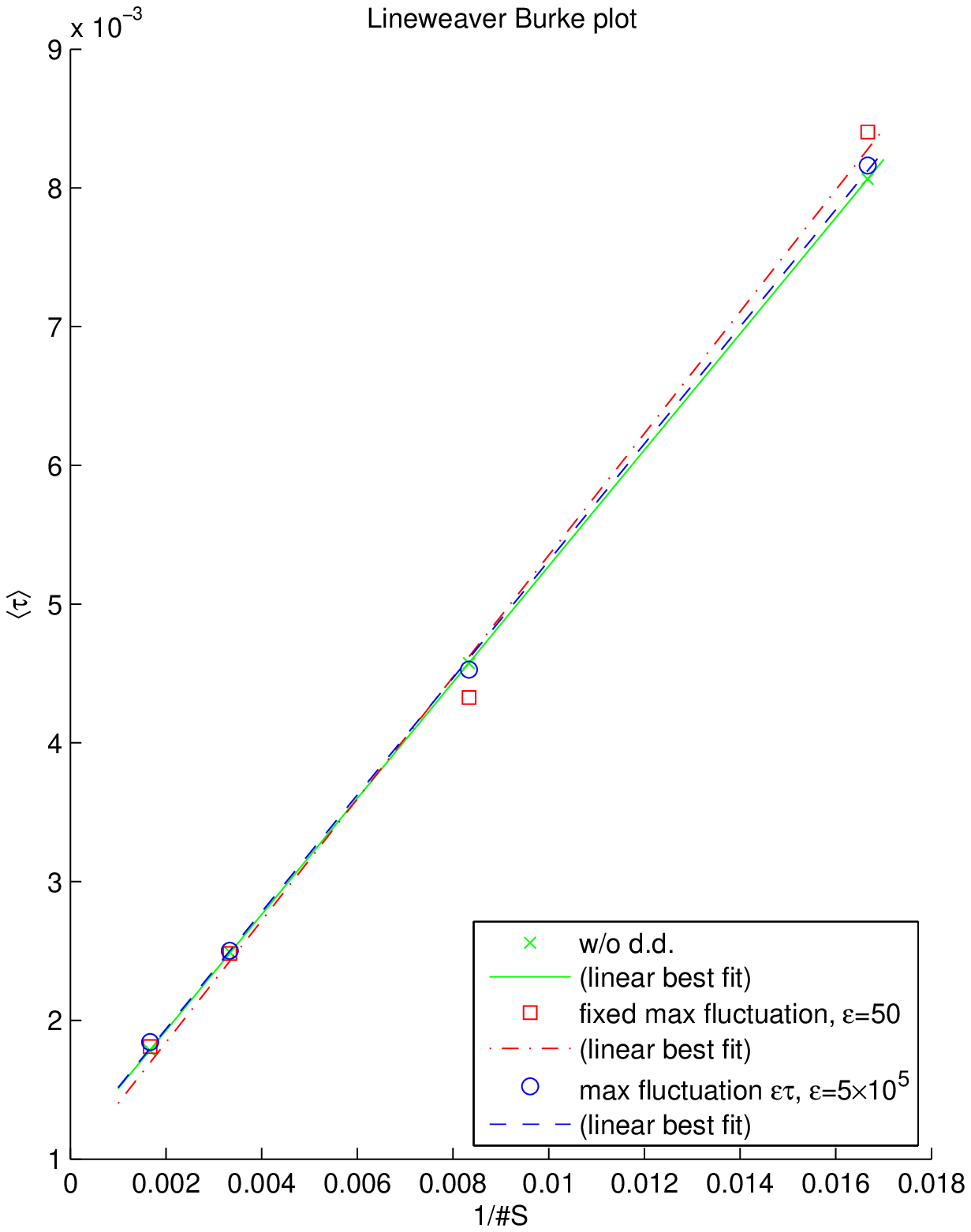}
\caption{The Lineweaver-Burke plot shows the linear relationship between the inverse of the substrate amount and the mean turnover waiting time, for the cases with and without dynamic disorder.}
\label{Fig: Lineweaver-Burke plots}
\end{figure}

As suggested by equations (\ref{Eqn: single enzyme MM}) and (\ref{Eqn: dynamic disorder MM}), the Lineweaver-Burke plots (Fig. \ref{Fig: Lineweaver-Burke plots}) obtained for the models with and without dynamic disorder reveal the linear relationship between the mean turnover time $\langle\tau\rangle$ and the inverse substrate concentrations. The slope and $y$-intercept for the linear best fits for this data are shown in the following table. Also shown are the estimated values of $\chi$ and $C_M$ for the models with dynamic disorder, and true and estimated values of $\bar{c}$ and $K_M$ for the model without dynamic disorder. 
The time-dependent $c$-dynamics model gives a more accurate estimate of $C_M$, though the estimate of $\chi$ is still not accurate.

\begin{table}
\begin{tabular}{|c|c|c|c|}
\hline
    & Fixed $\varepsilon$ model & Time-dependent     & W/o dynamic \\
    &                           & $c$-dynamics model & disorder \\\hline
  Slope     & 0.4389 & 0.4219 & 0.41861 \\
  $y$-intercept & 0.0009637 & 0.001095 & 0.001088 \\\hline
  Estimated $\chi$ or $\bar{c}$ & 1037.7 & 913.5 & 918.8 \\
  Estimates $C_M$ or $K_M$  & 455.4 & 385.4 & 384.9 \\\hline
  True $\chi$ or $\bar{c}$  & 704 & 704 & 924 \\
  True $C_M$ or $K_M$ & 380.1 & 380.1 & 384.5\\\hline
\end{tabular}
\end{table}

\subsection{Computational efficiency.}

The modified algorithm requires four random number draws per reaction step, as compared to two for the original Gillespie algorithm. Given the added complexity of the model, this is not an excessively large computational burden.
Also, the modified algorithm suffers from the same issue of computational efficiency that the original Gillespie algorithm faces, due to have to evolve the system reaction by reaction.
Techniques to speed up the computing time, such as the \emph{next reaction method}, can be applied to improve the computational efficiency in this modification of the algorithm, and will be the subject of future work.
However, the largest computational cost arises from having to invert equation \eqref{Eqn: waiting time in system} to find $\tau$. 
Approximations of $\tau$, like the linearization done in Section \eqref{Section: Time dependent c-dynamics}, may be employed if the error can be properly quantified.

\section{Discussion and Conclusion}

We have provided a framework for the modified Gillespie algorithm to address the problem of stochastic simulation of biochemical systems possessing dynamic disorder. 
Although the modelling and implementation shown in the single enzyme reaction examples leave much room for refinement, it nonetheless is able to pinpoint some concrete modelling ingredients that are corroborated by experimental data, and is a versatile method that can be adapted to many model dynamics.
This gives us a good indication of the direction that further modelling efforts can take. 
With this framework, model calibration against real data will be possible.

%%%%%%%%%%%%%%%%%%%%%%%%%%%%%%%%%%%%

\appendix 
\section{Derivation of $f(\tau|c_0)$}\label{Appdx: tau, c0}
In this appendix, we derive the formula for the conditional probability $f(\tau|c_0)$, assuming that we have available a model for the transition kernel $f_{c_1|\tau,c_0}(c_1|\tau,c_0)$ for the process $c(t)$.

Suppose that at time $t$ the rate constant is $c_0$. 
Let $P_0(\tau|c_0)$ be the conditional probability that \emph{no} reaction occurs within the next $\tau$ time.
Then the conditional probability that the next reaction occurs in the infinitesimal time interval $[t+\tau, t+\tau+\delta\tau)$ is approximately $f_{\tau|c_0}(\tau|c_0) \delta \tau$, and
\begin{align*}
  &f_{\tau|c_0}(\tau|c_0) \delta \tau\\
  &= P_0(\tau|c_0) \times Pr(\text{reaction occurs in } [t+\tau,t+\tau+\delta\tau) | c_0)
\end{align*}
For an arbitrary $\tau$, recalling the transition kernel $f_{c_1|\tau,c_0}(c_1|\tau,c_0)$ for the underlying process $c(t)$, we condition on the value of $c_{\tau}=c(t+\tau)$ at time $t+\tau$,
\begin{align*}
  &Pr(\text{reaction occurs in } [t+\tau,t+\tau+\delta\tau) | c_0) \\
  &= \int_{\mathbb{R}^{+}} Pr(\text{reaction occurs in } [t+\tau,t+\tau+\delta\tau) | c_{\tau}, c_0) \cdot f_{c_\tau|\tau,c_0}(c_{\tau}|\tau,c_0) dc_{\tau} \\
  &= \int_{\mathbb{R}^{+}} h c_{\tau} \delta\tau f_{c_\tau|\tau,c_0}(c_{\tau}|\tau,c_0) dc_{\tau} \\
  &= h \varphi_{c_0}(\tau) \delta\tau
\end{align*}
where $\varphi_{c_0}(\tau)=\mathbb{E}_{c_1}(c_1|\tau,c_0)$. 
In the event that the reaction occurs in $ [t+\tau,t+\tau+\delta\tau)$, we will have that $c_1 = c_{\tau}$.

To find $P_0(\tau|c_0)$, 
\begin{align*}
  &P_0(\tau + \delta\tau|c_0) \\
  &= P_0(\tau|c_0) \times Pr(\text{No reaction occurs in } [t+\tau,t+\tau+\delta\tau) | c_0) \\
  &=  P_0(\tau|c_0) \times \left(1- h \varphi_{c_0}(\tau) \delta\tau \right)
\end{align*}
So $P_0(\tau|c_0)$ satisfies a differential equation
\begin{align*}
  \frac{d \log P_0}{d\tau} = - h \varphi_{c_0}(\tau)
\end{align*}
with the initial condition $P_0(\tau=0|c_0)=1$. 
Hence, $P_0(\tau|c_0) = e^{-h \int_0^{\tau} \varphi_{c_0}(\tau') d\tau'}$, and
\begin{align*}
  f_{\tau|c_0}(\tau|c_0) = h \varphi_{c_0}(\tau) e^{-h \int_0^{\tau} \varphi_{c_0}(\tau') d\tau'}.
\end{align*}

\end{document}